\documentclass[pra, 12 pt]{revtex4}
\usepackage[english]{babel}
\usepackage{amsmath}
\usepackage{epsfig}

\begin{document}
\title{Anomalous Diffusion in Velocity Space}
\author{S.A. Trigger}
\address{Joint\, Institute\, for\, High\, Temperatures, Russian\, Academy\,
of\, Sciences, 13/19, Izhorskaia Str., Moscow\, 127412, Russia;\\
email:\,satron@mail.ru}

\begin{abstract}

The problem of anomalous diffusion in the momentum space is
considered on the basis of the appropriate probability transition
function (PTF). New general equation for description of the
diffusion of heavy particles in the gas of the light particles is
formulated on basis of the new approach similar to one in
coordinate space [1]. The obtained results permit to describe the
various situations when the probability transition function (PTF)
has a long tail in the momentum space. The effective friction and
diffusion coefficients are found.\\

PACS number(s): 52.27.Lw, 52.20.Hv, 05.40.-a, 05.40.Fb

\end{abstract}

\maketitle

\section{Introduction}
Interest in anomalous diffusion is conditioned by a large variety
of applications: semiconductors, polymers, some granular systems,
plasmas in specific conditions, various objects in biological
systems, physical-chemical systems, et cetera.

The deviation from the linear in time $<r^2(t)>\sim t$ dependence
of the mean square displacement have been experimentally observed,
in particular, under essentially non-equilibrium conditions or for
some disordered systems. The average square separation of a pair
of particles passively moving in a turbulent flow grows, according
to Richardson's law, with the third power of time [2]. For
diffusion typical for glasses and related complex systems [3] the
observed time dependence is slower than linear. These two types of
anomalous diffusion obviously are characterized as superdiffusion
$<r^2(t)>\sim t^\alpha$ $(\alpha>1)$ and subdiffusion $(\alpha<1)$
[4]. For a description of these two diffusion regimes a number of
effective models and methods have been suggested. The continuous
time random walk (CTRW) model of Scher and Montroll [5], leading
to strongly subdiffusion behavior, provides a basis for
understanding photoconductivity in strongly disordered and glassy
semiconductors. The Levy-flight model [6], leading to
superdiffusion, describes various phenomena as self-diffusion in
micelle systems [7], reaction and transport in polymer systems [8]
and is applicable even to the stochastic description of financial
market indices [9]. For both cases the so-called fractional
differential equations in coordinate and time spaces are applied
as an effective approach [10].

However, recently a more general approach has been suggested in
[1], [11], which avoid the fractional differentiation, reproduce
the results of the standard fractional differentiation method,
when the last one is applicable, and permit to describe the more
complicated cases of anomalous diffusion processes. In [12] these
approach has been applied also to the diffusion in the
time-dependent external field.

In this paper the problem of anomalous diffusion in the momentum
(velocity) space will be considered. In spite of formal
similarity, diffusion in the momentum space is very different
physically from the coordinate space diffusion. It is clear
already because the momentum conservation, which take place in the
momentum space has no analogy in the coordinate space.

The anomalous diffusion in the velocity space is weakly
investigated. Some attempts to investigate the influence of the
long tails of correlation functions in velocity space have been
done recently by W. Ebeling and M.Yu. Romanovsky (Contr. Plasma
Physics, in print, private communication). The consequent way to
describe the anomalous diffusion in the velocity space is,
according to our knowledge, still absent.

In Section II the diffusion equation in coordinate space for a
homogeneous system is shortly reviewed. The diffusion in velocity
space for the cases of normal and anomalous behavior of the
probability transition function PTF is presented in Sections III,
IV.

\section{Diffusion in the coordinate space on the basis of
a master-type equation}

Let us consider diffusion in coordinate space on the basis of the
master equation, which describes the balance of grains coming in
and out the point $r$ at the moment $t$. The structure of this
equation is formally similar to the master equation in the
momentum space (see, e.g., [1],[11]). Of course, for coordinate
space there is no conservation law, similar to that in momentum
space:
\begin{equation}
\frac{df_g({\bf r},t)}{dt} = \int d{\bf r'} \left\{W ({\bf r, r'})
f_g({\bf r', t}) - W ({\bf r', r}) f_g({\bf r},t) \right\}.
\label{DC1}
\end{equation}
The probability transition $W({\bf r, r'})$ describes the
probability for a grain to transfer from the point ${\bf r'}$ to
the point ${\bf r}$ per unit time. We can rewrite this equation in
the coordinates ${\bf \rho= r'- r}$ and ${\bf r}$ as:
\begin{eqnarray}
\frac{df_g({\bf r},t)}{dt} = \int d{\bf \rho} \left\{W ({\bf \rho,
r+\rho}) f_g({\bf r+\rho},t) - W ({\bf \rho, r}) f_g({\bf r},t)
\right\}. \label{DC2}
\end{eqnarray}
Assuming that the characteristic displacements are small one may
expand Eq.~(\ref{DC2}) and arrive at the Fokker-Planck form of the
equation for the density distribution $f_g({\bf r},t)$
\begin{equation}
\frac{df_g({\bf r},t)}{dt} = \frac {\partial}{\partial r_{\alpha}}
\left[A_\alpha ({\bf r}) f_g({\bf r},t) + \frac {\partial}
{\partial r_{\beta}} \left( B_{\alpha\beta}({\bf r}) f_g({\bf
r},t) \right)\right]. \label{DC3}
\end{equation}
The coefficients $A_\alpha$ and $B_{\alpha \beta}$, describing the
acting force and diffusion, respectively, can be written as
functionals of the PTF in the coordinate space $W$ (with the
dimension $s$) in the form:
\begin{equation}
A_\alpha({\bf r}) = \int d^s \rho \rho_\alpha W({\bf \rho, r})
\label{DC4}
\end{equation}
and
\begin{equation}
B_{\alpha\beta}({\bf r})= \frac{1}{2}\int d^s\rho \rho_\alpha
\rho_\beta W({\bf\rho, r}). \label{DC5}
\end{equation}
For the isotropic case the probability function depends on ${\bf
r}$ and the modulus of $\rho$. For a homogeneous medium, when
$r$-dependence of the PT is absent, the coefficients $A_\alpha=0$
while the diffusion coefficient is constant with
$B_{\alpha\beta}=\delta_{\alpha\beta}B$, where B is the integral
\begin{equation}
B = \frac{1}{2s}\int d^s \rho \rho^2 W(\rho). \label{DC6}
\end{equation}

This consideration cannot be applied to specific situations in
which the integral in Eq.~(\ref{DC6}) is infinite. In that case we
have to examine the general transport equation (\ref{DC1}). We
will now consider the problem for the homogeneous and isotropic
case, when the PT function depends only on $|\rho|$. By
Fourier-transformation we arrive at the following form [11] of
Eq.~(\ref{DC1}):
\begin{equation}
\frac{df_g({\bf k},t)}{dt} = \int d^s \rho  \left[\exp (i {\bf
k}\rho) -1 \right]W(|\rho|) f_g({\bf k},t)\equiv X({\bf
k})f_g({\bf k},t), \label{DC7}
\end{equation}
where $X({\bf k})\equiv X(k)$. Let us assume a simple form of the
PT function with a power dependence on the distance $W(\rho)=
C/|\rho|^\alpha$, where $C$ is a constant and $\alpha>0$. Such
type singular dependence is typical for jump diffusion probability
in heteropolymers in solution (see, e.g., [13], where the
different applications of anomalous diffusion are considered on
the basis of the fractional differentiation method). For the
one-dimensional case we find:
\begin{equation}
X(k)\equiv -4 \int_0^\infty d u \, sin^2
\left(\frac{k\,u}{2}\right)W(u)= - 2^{3-\alpha}C |k|^{\alpha-1}
\int_0^\infty d\zeta \frac{sin^2\zeta}{\zeta^\alpha}. \label{DC8}
\end{equation}
For the values $1<\alpha<3$ this function is finite and equal to
\begin{equation}
X(k) = - \frac{C\, \Gamma[(3-\alpha)/2]\,|k|^{\alpha-1}}{2^\alpha
\sqrt\pi \, \Gamma(\alpha/2)(\alpha-1)}  , \label{DC9}
\end{equation}
where $\Gamma$ is the Gamma-function. At the same time the
integral in Eq.~(\ref{DC6}) for  such a type of PT functions is
infinite, because usual diffusion is absent.

At the same time the integral (\ref{DC6}) for  such a type of PT
functions is infinite, because usual diffusion is absent. The
procedure considered for the simplest cases of power dependence of
the PT function is equivalent to the equation with fractional
space differentiation [10],[13]:
\begin{equation}
\frac{df_g(x,t)}{dt} = C\Delta^{\mu/2}f_g(x,t),\label{DC10}
\end{equation}
where $\Delta^{\mu/2}$ is a fractional Laplacian, a linear
operator, whose action on the function $f(x)$ in Fourier space is
described by $\Delta^{\mu/2}f(x)=-(k^2)^{\mu/2}f(k)=-|k|^\mu
f(k)$. In the case considered above $\mu\equiv(\alpha-1)$, where
$0<\mu<2$. For more general PT functions, which (for arbitrary
values $\rho$) are not proportional to the $\alpha$ power of
$\rho$, the method described above is also applicable, although
the fractional derivative does not exist.

For the case of purely power dependence of PT the non-stationary
solution for the density distribution describes so-called
super-diffusion (or Levy flights). The solution of
Eq.~(\ref{DC10}) in Fourier space reads:
\begin{equation}
f_g(k,t) = \exp (-C|k|^\mu t),\label{DC11}
\end{equation}
which in coordinate space corresponds to a so-called symmetric
Levy stable distribution:
\begin{equation}
f_g(x,t) = \frac {1}{(kt)^{1/\mu}} L\left[\frac{x}{(kt)^{1/\mu}};
\mu, 0 \right]. \label{DC12}
\end{equation}
For the general case it follows from Eq.~(\ref{DC7}) that
\begin{equation}
f_g(k,t) = C_1 \exp [X(k)t],\label{DC12a}
\end{equation}
with some constant $C_1$.

The consideration on the basis of PTF function given above,
permits to avoid the fractional differentiation method and to
consider more general physical situations of the non-power
probability transitions for arbitrary space dimension.

\section{Diffusion in the velocity space on the basis of
a master-type equation}

Let us consider now the main problem formulated in the
introduction, namely, diffusion in velocity space ($V$-space) on
the basis of the respective master equation, which describes the
balance of grains coming in and out the point $p$ at the moment
$t$. The structure of this equation is formally similar to the
master equation in the coordinate space Eq.~(\ref{DC2})
\begin{equation}
\frac{df_g({\bf p},t)}{dt} = \int d{\bf q} \left\{W ({\bf q, p+q})
f_g({\bf p+q, t}) - W ({\bf q, p}) f_g({\bf p},t) \right\}.
\label{DC2b}
\end{equation}
Of course, for coordinate space there is no conservation law,
similar to that in the momentum space. The probability transition
$W({\bf p, p'})$ describes the probability for a grain with
momentum ${\bf p'}$ (point ${\bf p'}$) to transfer from this point
${\bf p'}$ to the point ${\bf p}$ per unit time. The momentum
transferring is equal ${\bf q= p'- p}$. Assuming in the beginning
that the characteristic displacements are small one may expand
Eq.~(\ref{DC2})  and arrive at the Fokker-Planck form of the
equation for the density distribution $f_g({\bf p},t)$
\begin{equation}
\frac{df_g({\bf p},t)}{dt} = \frac {\partial}{\partial p_\alpha}
\left[ A_\alpha ({\bf p}) f_g({\bf p},t) + \frac{\partial}
{\partial p_\beta} \left(B_{\alpha\beta}({\bf p}) f_g({\bf p},t)
\right)\right]. \label{DC3b}
\end{equation}
\begin{equation}
A_\alpha({\bf p}) = \int d^s q q_\alpha W({\bf q, p});\;\;\;\
B_{\alpha\beta}({\bf p})= \frac{1}{2}\int d^s q q_\alpha q_\beta
W({\bf q, p}). \label{DC4b}
\end{equation}
The coefficients $A_\alpha$ and $B_{\alpha \beta}$ describing the
friction force and diffusion, respectively.

Because the velocity of heavy particles is small, the $\bf
p$-dependence of the PTF can be neglected for calculation of the
diffusion, which in this case is constant
$B_{\alpha\beta}=\delta_{\alpha\beta}B$, where B is the integral
\begin{equation}
B = \frac{1}{2s}\int d^s q q^2 W(q). \label{DC6b}
\end{equation}

If to neglect the $\bf p$-dependence of the PTF at all we arrive
to the coefficient $A_\alpha=0$ (while the diffusion coefficient
is constant). This neglecting, as well known is wrong, and the
coefficient $A_\alpha$ for the Fokker-Planck equation can be
determined by use the argument that the stationary distribution
function is Maxwellian. On this way we arrive to the standard form
of the coefficient $MT A_\alpha(p)=p_\alpha B$, which is one of
the forms of Einstein relation. For the systems far from
equilibrium this argument is not acceptable.

To find the coefficients in the kinetic equation, which are
applicable also to slowly decreasing PT functions, let us use a
more general way, based on the difference of the velocities of the
light and heavy particles. For calculation of the function
$A_\alpha$ we have take into account that the function $W(\bf
{q,p})$ is scalar and depends on $q, {\bf q \cdot p}, p$.
Expanding $W(\bf {q,p})$ on $\bf {q\cdot p}$ one arrive to the
approximate representation of the functions $W(\bf {q,p})$ and
$W({\bf q, p+q})$:
\begin{eqnarray}
W({\bf q,p)}\simeq W(q)+\tilde W'(q)({\bf q \cdot p})+
 \frac{1}{2}\tilde W''(q) ({\bf q \cdot p})^2 . \label{DC7b}
\end{eqnarray}
\begin{equation}
W ({\bf q, p+q})\simeq W(q)+\tilde W'(q) \,({\bf q \cdot
p})+\frac{1}{2}\tilde W''(q) ({\bf q \cdot p})^2 +q^2\tilde
W'(q),\label{DC9b}
\end{equation}
where $\tilde W'(q)\equiv \partial W (q, {\bf q \cdot p})
/\partial ({\bf q p})\mid_{{\bf q \cdot p}=0}$ and $\tilde
W''(q)\equiv \partial^2 W (q, {\bf q \cdot p}) /\partial ({\bf q
p})^2 \mid_{{\bf q \cdot p}=0}$.

Then, with the necessary accuracy, $A_\alpha$ equals
\begin{equation}
A_\alpha({\bf p}) = \int d^s q q_\alpha q_\beta p_\beta \tilde
W'(q)= p_\alpha \int d^s q q_\alpha q_\alpha  \tilde
W'(q)=\frac{p_\alpha}{s} \int d^s q q^2  \tilde W'(q)\label{DC10b}
\end{equation}
If for the function $W({\bf q,p)}$ the equality $\tilde W'(q)=
W(q)/ 2 MT$ is fulfilled, then we arrive to the usual Einstein
relation
\begin{equation}
M T A_\alpha({\bf p}) =  p_\alpha B \label{DC11b}
\end{equation}

Let us check this relation for the Boltzmann collisions, which are
described by the PT-function $W({\bf q, p)}=w_B({\bf q, p})$ [11]:
\begin{eqnarray}
w_B({\bf q, p})=\frac{2\pi}{\mu^2 q} \int_{q/2\mu}^\infty du\,u\,
\frac{d \sigma}{do} \left[\arccos \, (1-\frac{q^2}{2\mu^2 u^2}), u
\right] f_b (u^2+ v^2-{\bf q \cdot v} /\mu), \label{DC12b}
\end{eqnarray}
where (${\bf p}=M{\bf v}$) and $d \sigma / do$ and $f_b$ are
respectively the differential cross-section for scattering and the
distribution function for the light particles. For the equilibrium
Maxvellian distribution $f_b^0$ the equality $\tilde W'(q)= W(q)/
2 MT$ is evident and we arrive to the usual Fokker-Planck equation
in velocity space with the constant diffusion $D \equiv B /M^2$
and friction $\beta \equiv B/MT=DM/T$ coefficients, which satisfy
the Einstein relation.

For some non-equilibrium situations the PTF can possess a long
tail. In this case we have derive a generalization of the
Fokker-Planck equation in spirit of the above consideration for
the coordinate case, because the diffusion and friction
coefficients in the form Eqs.~(\ref{DC6b}),(\ref{DC10b}) diverge
for large $q$ if the functions have the asymptotic behavior
$W(q)\sim 1/q^\alpha$ with $\alpha\leq s+2$ and (or) $\tilde
W'(q)\sim 1/q^\beta$ with $\beta \leq s+2$.

Let us insert in Eq.~(\ref{DC2b}) the expansions for $W$ (as an
example we choose $s=3$, the arbitrary $s$ can be considered by
the similar way). With necessary accuracy we find
\begin{eqnarray}
\frac{df_g({\bf p},t)}{dt}=\int d{\bf q} \{f_g({\bf p+q}, t)[1+
q_\alpha \partial /\partial
p_\alpha][W(q)+\tilde W'(q) \,({\bf q \cdot p})+\nonumber\\
\frac{1}{2}\tilde W''(q) ({\bf q \cdot p})^2 ]-f_g({\bf
p},t)[W(q)+\tilde W'(q) \,({\bf q \cdot p})+\frac{1}{2}\tilde
W''(q) ({\bf q \cdot p})^2]\} \label{DC13b}
\end{eqnarray}

After the Fourier-transformation $f({\bf r})=\int \frac{d{\bf
p}}{(2\pi)^3} exp(i{\bf p r})f({\bf p},t)$ Eq.~(\ref{DC13b})
reads:
\begin{eqnarray}
\frac{df_g({\bf r},t)}{dt} = \int d{\bf q} \{ exp(-i{\bf(q
r)}[W(q)- i \tilde W'(q) \,({\bf
q} \cdot \frac{\partial}{\partial {\bf r}}) +\nonumber\\
-\frac{1}{2}\tilde W''(q) ({\bf q} \cdot \frac{\partial}{\partial
{\bf r}})^2] - [W(q) - i \tilde W'(q) \,({\bf q} \cdot
\frac{\partial}{\partial {\bf r}})-\frac{1}{2}\tilde W''(q) ({\bf
q} \cdot\frac{\partial}{\partial {\bf r}})^2]\}f_g({\bf r},
t)\label{DC15b}
\end{eqnarray}

We can rewrite this equation as
\begin{eqnarray}
\frac{df_g({\bf r},t)}{dt} = A(r)f({\bf r})+ B_\alpha
(r)\frac{\partial}{\partial {\bf r}_\alpha} f({\bf
r},t)+C_{\alpha\beta}(r)\frac{\partial^2}{\partial {\bf r}_\alpha
\partial {\bf
r}_\beta}f({\bf r},t)\label{DC16b}
\end{eqnarray}
where
\begin{eqnarray}
A(r)= \int d{\bf q}  [exp(-i{\bf(q r) })-1]W(q)= 4\pi
\int_0^\infty dq q^2 \left[\frac{sin\, (q r)}{qr}-1\right]W(q)
\label{DC17b}
\end{eqnarray}
\begin{eqnarray}
B_\alpha\equiv r_\alpha B(r);\;B(r)=-\frac{i}{r^2} \int d{\bf q}
{\bf q r} [exp(-i{\bf(q r})-1]  \tilde W'(q)= \nonumber\\
\frac{4\pi}{r^2} \int_0^\infty dq q^2 \left[cos\, (q r)-\frac{sin
(q r)}{q r}\right]W'(q) \label{DC19b}
\end{eqnarray}
\begin{eqnarray}
C_{\alpha\beta} (r)\equiv r_\alpha r_\beta C(r)= -\frac{1}{2}\int
d{\bf q} q_\alpha q_\beta [exp(-i{\bf(q r})-1]  \tilde
W''(q)\label{DC20b}
\end{eqnarray}
\begin{eqnarray}
C(r )=-\frac{1}{2 r^4} \int d{\bf q} {\bf (q r)^2} [exp(-i({\bf q
r})-1] \tilde W''(q)=\nonumber\\\frac{2\pi}{r^2} \int_0^\infty dq
q^4 \left[\frac{2 sin (q r)}{q^3 r^3}-\frac{2 cos\, (q r)}{q^2
r^2}-\frac{sin (qr)}{qr}+\frac{1}{3}\right]W''(q) \label{DC21b}
\end{eqnarray}

For the isotropic function $f({\bf r})=f(r)$ one can rewrite
Eq.~(\ref{DC16b}) in the form
\begin{eqnarray}
\frac{df_g(r,t)}{dt} = A (r) f(r)+ B(r)r \frac{\partial}{\partial
r} f(r)+C(r) r^2 \frac{\partial^2}{\partial r^2} f(r)\label{DC22b}
\end{eqnarray}

For the case of strongly decreasing PDF the exponent under the
integrals for the functions $A(r)$, $B(r)$ and $C(r)$ can be
expanded
\begin{eqnarray}
A(r)\simeq=-\frac{r^2}{6}\int d{\bf q}\, q^2 W(q);\; B(r)\simeq -
\frac{1}{6} \int d{\bf q} \, q^2 \tilde W'(q);\;C(r )\simeq 0.
\label{DC23b}
\end{eqnarray}

Then the simplified kinetic equation for the case of short-range
on $q$-variable PTF (non-equilibrium, in general case) reads
\begin{eqnarray}
\frac{df_g(r,t)}{dt} = A_0 r^2 f(r)+ B_0 r
\frac{\partial}{\partial r} f(r),\label{DC25b}
\end{eqnarray}
where $A_0\equiv -1/6 \int d{\bf q}\, q^2 W(q)$ and $B_0\equiv
-1/6 \int d{\bf q} \, q^2 \tilde W'(q)$.

Stationary solution of Eq.~(\ref{DC22b}) for $C(r)=0$ reads
\begin{eqnarray}
f_g(r,t)=C exp\,\left[-\int_0^r dr'\frac{A(r')}{r' B(r')}\right]
=C exp\,\left[-\frac{A_0 r^2}{2 B_0}\right] \label{DC26b}
\end{eqnarray}
The respective normalized stationary momentum distribution equals
\begin{eqnarray}
f_g(p)=\frac{N_g B_0^{3/2}}{(2 \pi A_0)^{3/2}}exp \,[-\frac{B_0
p^2}{2 A_0}]\label{DC28b}
\end{eqnarray}
Therefore in  Eq.~(\ref{DC23b}) the constant $C=N_g$. Equation
(\ref{DC25b}) and this distribution are the generalization of the
Fokker-Planck case for normal diffusion on non-equilibrium
situation, when the prescribed $ W({\bf q,p})$ is determined,
e.g., by some non-Maxwellian distribution of the small particles
$f_b$. To show this by other way let us make the Fourier
transformation of (\ref{DC16b}) with $C=0$ and the respective $A$
and $B_\alpha$:
\begin{eqnarray}
\frac{df_g({\bf p},t)}{dt} = - A_0 \frac{\partial^2}{\partial p^2}
f_g({\bf p},t)- B_0 \frac{\partial}{\partial p_\alpha}p_\alpha f_g
({\bf p},t),\label{DC29b}
\end{eqnarray}
Therefore we arrive to the Fokker-Planck type equation with the
friction coefficient $\beta\equiv-B_0$ and diffusion coefficient
$D=-A_0/M^2$. In general these coefficients (Eq.~(\ref{DC23b})) do
not satisfy to the Einstein relation.

In the case of equilibrium $W$-function (e.g., $f_b=f_b^0$, see
above) the equality $\tilde W'(q)=W(q)/2M T_b$ is fulfilled. Then
$A(r)/r B(r)=M T_b r$ ($A_0=MT_b B_0$). Only in this case the
Einstein relation between the diffusion and friction coefficients
exists and the standard Fokker-Planck equation is valid.

\section{The model of anomalous diffusion in $V$ - space}

Now we can calculate the coefficients for the models of anomalous
diffusion.

In this paper we calculate only the simple model system of the
hard spheres with the different masses $m$ and $M\gg m$,
$d\sigma/do=a^2/4$. Let us suppose that in the model under
consideration the small particles are described by the prescribed
stationary distribution $f_b=n_b \phi_b /u_0^3$ (where $\phi_b$ is
non-dimensional distribution, $u_0$ is the characteristic velocity
for the distribution of the small particles) and $\xi \equiv (u^2+
v^2-{\bf q\cdot v} /\mu)/u_0^2$.
\begin{eqnarray}
W_a ({\bf q, p})= \frac{n_b a^2 \pi}{2\mu^2 u_0 q}
\int^\infty_{(q^2/4\mu^2+v^2-{\bf{q\cdot v}}/\mu)/u_0^2} d\xi
\,\cdot \phi _b (\xi).\label{DC30b}
\end{eqnarray}

If the distribution $\phi_b(\xi)=1/\xi^\gamma$ ($\gamma>1$)
possess a long-tail we get
\begin{eqnarray}
W_a({\bf q, p})= \frac{n_b a^2 \pi}{2\mu^2 u_0
q}\frac{\xi^{1-\gamma}}{(1-\gamma)}|_{\xi_0}^\infty= \frac{n_b a^2
\pi}{2\mu^2 u_0 q}\frac{\xi_0^{1-\gamma}}{(\gamma-1)}
,\label{DC31b}
\end{eqnarray}
where $\xi_0\equiv (q^2/4\mu^2+v^2-{\bf{q\cdot v}}/\mu)/u_0^2$.

For the case $p=0$ the value $\xi_0 \rightarrow \tilde \xi_0
\equiv q^2/4\mu^2 u_0^2$ and we arrive to the expression for
anomalous $W \equiv W_a$
\begin{eqnarray}
W_a ({\bf q, p=0})=\frac{n_b a^2
\pi}{2^{3-2\gamma}(\gamma-1)\mu^{4-2\gamma}
u_0^{3-2\gamma}q^{2\gamma-1}}\equiv \frac{C_a}{q^{2\gamma-1}}
.\label{DC32b}
\end{eqnarray}

The function $A(r)$, according to Eq.~(\ref{DC17b})
\begin{eqnarray}
A(r)\equiv 4\pi \int_0^\infty dq q^2 \left[\frac{sin\, (q
r)}{qr}-1\right]W(q)= 4\pi C_a
 \int_0^\infty d q
\frac{1}{q^{2\gamma-3}} \left[\frac {sin(qr)}{qr}-1\right]
\label{DC33b}
\end{eqnarray}
Comparing the reduced equation (see below) in the velocity space
with the diffusion in coordinate space
($2\gamma-1\leftrightarrow\alpha$ and $W(q)=C/q^{2\gamma-1}$) we
can establish that the convergence of the integral in the right
side of Eq.~(\ref{DC33b}) (3d case) is provided if $3<2\gamma-1<5$
or $2<\gamma<3$. The inequality $\gamma<3$ provides the
convergence for small $q$ ($q\rightarrow0$) and the inequality
$\gamma>2$ provides the convergence for $q\rightarrow\infty$.

Now to determine the structure of the transport process and the
kinetic equation in the velocity space we have find the functions
$\tilde W'(q)$ and $\tilde W''(q)$.

If $p\neq 0$ to find $\tilde W'(q)$ and $\tilde W''(q)$ we have
use the full value $\xi_0\equiv (q^2/4\mu^2+p^2/M^2-{\bf{q\cdot
p}}/M \mu)/u_0^2$ and it derivatives on ${\bf q \cdot p}$ at
$p=0$, $\xi'_0=-1/M \mu u_0^2$ and $\xi''_0=0$. Then
\begin{eqnarray}
\tilde W'({\bf q, p})\equiv \frac{n_b a^2 \pi}{2M \mu^3 u^3_0
q}\xi_0^{-\gamma};
 \;\; \; \tilde W''({\bf q, p})\equiv \frac{n_b
a^2 \pi\gamma}{2M^2 \mu^4 u^5_0 q}\xi_0^{-\gamma-1} \label{DC34b}
\end{eqnarray}
Therefore for $p=0$ ($\xi_0 \rightarrow \tilde \xi_0$) we obtain
the functions
\begin{eqnarray}
\tilde W'(q)\equiv \frac{(4 \mu^2u_0^2)^\gamma n_b a^2 \pi}{2M
\mu^3 u^3_0 q^{2\gamma+1}};
 \;\; \; \tilde W''(q)\equiv \frac{(4 \mu^2u_0^2)^{\gamma+1} n_b
a^2 \pi\gamma}{2M^2 \mu^4 u^5_0 q^{2\gamma+3}} \label{DC35b}
\end{eqnarray}
We have establish now the conditions of convergence the integrals
for $B(r)$ and $C(r)$.

\begin{eqnarray}
B(r)= \frac{4\pi}{r^2} \int_0^\infty dq q^2 \left[cos\, (q
r)-\frac{sin (q r)}{q r}\right]W'(q) \label{DC36b}
\end{eqnarray}

Convergence $B(r)$ exists for small $q$ if $\gamma<2$ and for
large $q\rightarrow\infty$ for $\gamma>1/2$.

Finally for $C(r)$ convergence is determined by the equalities
$\gamma<2$ for small $q$ and $\gamma>1$ for large $q$
\begin{eqnarray}
C(r )=\frac{2\pi}{r^2} \int_0^\infty dq q^4 \left[\frac{2 sin (q
r)}{q^3 r^3}-\frac{2 cos\, (q r)}{q^2 r^2}-\frac{sin
(qr)}{qr}+\frac{1}{3}\right]W''(q) \label{DC37b}
\end{eqnarray}

Therefore to provide convergence for $A$, $B$, $C$ for large $q$
we have provide convergence for $A$, that means $\gamma>2$. To
provide convergence for small $q$ enough to provide convergence
for $B$ and $C$, that means $\gamma<2$. Therefore for the purely
power behavior of the function $f_b(\xi)$ convergence is absent.
However, for existence of the anomalous diffusion in the momentum
space in reality the convergence for small $q$ is always provided,
e.g. by finite value of $v$ or by change of the small $q$-behavior
of $W(q)$ (compare with the examples of anomalous diffusion in
coordinate space [1]). Therefore the "anomalous diffusion in
velocity space"  for the power behavior of $W(q)$, $W'(q)$ and
$W''(q)$ on large $q$ exists if for large $q$ the asymptotic
behavior of $W(q\rightarrow\infty)\sim 1/q^{2\gamma-1}$ with
$\gamma>2$. At the same time the expansion of the exponential
function in Eqs. (\ref{DC17b})-(\ref{DC21b}) under the integrals,
which leads to the Fokker-Planck type kinetic equation is invalid
for the power-type kernels $W(\bf {q, p)}$.

\section{Conclusions}
In the previous sections we shortly reviewed the anomalous
diffusion in the coordinate space and firstly consequently
considered the problem of anomalous diffusion in momentum
(velocity) space. The new kinetic equation for anomalous diffusion
in velocity space is established. For the normal diffusion the
friction and diffusion coefficient are found for the
non-equilibrium case. For equilibrium case the usual Fokker-Plance
equation is reproduced as the particular case. The model of
anomalous diffusion in velocity space is described on the basis of
the respective expansion of the kernel in master equation and the
conditions of the convergence for the coefficients of the kinetic
equation are found.

\section*{Acknowledgment}
The author is thankful to  W. Ebeling, M. Yu. Romanovsky and I.M.
Sokolov for valuable discussions of some problems, reflected in
this work. I would like to thank M. S. Karavaeva for the permanent
stimulating interest. I also express my gratitude to the
Netherlands Organization for Scientific Research (NWO) and the
Russian Foundation for Basic Research (grant 07-02-01464-a) for
support of my investigations on the problems of stochastic
transport in gases, liquids and plasmas.


\begin{references}

\bibitem {1}
S.A. Trigger, G.J.F. van Heijst and P.P.J.M. Schram, {\it Journal
of Physics, Conference Series} {\bf 11}, 37 (2005).


\bibitem {2} Monin A S and Yaglom A M 1975 {\it Statistical Fluid
Mechanics: Mechanics of Turbulence} Vol II (Cambridge: MIT Press)

\bibitem {3} Rinn B,  Mass P and Bouchaud J P 2000 {\it Phys. Rev.
Lett.} {\bf 84}, 5405
\bibitem{4}
R. Metzler, and J. Klafter, {\it Phys. Rep.} \textbf{339}, 1
(2000).
\bibitem {5} Scher H and Montroll E W 1975 {\it Phys. Rev.} {\bf B12},
2455

\bibitem {6}  Klafter J,  Schlezinger M F and Zumofen G 1996 {\it Phys.
Today} {\bf 49}, 33

\bibitem {7} Ott A., Bouchaud J-P, Langevin D. and Urbach W. 1994 {\it
Phys. Rev. Lett.} {\bf 65}, 2201

\bibitem {8} Sokolov I.M. May J and Blumen A 1997 {\it Phys. Rev.
Lett.}\textbf{ 79}, 857

\bibitem {9} Gopikrishnan P, Plerou V, Amaral L A N, Meyer M and H.E.
Stanley H E 1999 {\it Phys. Rev.} {\bf E60}, 5305

\bibitem {10}
B.J. West, M. Bologna, and P.Grigolini, {\it Physics of Fractal
Operators}, (Springer-Verlag New York), 2003.

\bibitem {11}
S.A. Trigger, G.J.F. van Heijst and P.P.J.M. Schram, \emph{Physica
A} \textbf{347}, 77 (2005).

\bibitem {12}
S.A. Trigger, \emph{Physics Letters A} \textbf{372}, 8, p. 1288
(2008)

\bibitem {13}
D. Brockmann, and I.M. Sokolov, {\it Chemical Physics} {\bf 284},
409 (2002).



\end{references}
\end{document}